\documentclass[prd,aps,preprint,amsmath,nofootinbib,amssymb,eqsecnum,showkeys]{revtex4}
\usepackage{verbatim,graphics,graphicx,color,slashed}
\usepackage{ulem} % this one is for cross out text
\usepackage{hyperref}

\def\TeV{\textrm{TeV}}
\def\Tr{\textrm{Tr}}

\def\lsim{\mathrel {\vcenter {\baselineskip 0pt \kern 0pt
    \hbox{$<$} \kern 0pt \hbox{$\sim$} }}}
\def\gsim{\mathrel {\vcenter {\baselineskip 0pt \kern 0pt
    \hbox{$>$} \kern 0pt \hbox{$\sim$} }}}
\def\slashchar#1{\setbox0=\hbox{$#1$}           % set a box for #1
 \dimen0=\wd0                                 % and get its size
  \setbox1=\hbox{/} \dimen1=\wd1               % get size of /
\ifdim\dimen0>\dimen1                        % #1 is bigger
  \rlap{\hbox to \dimen0{\hfil/\hfil}}      % so center / in box
  #1                                        % and print #1
  \else                                        % / is bigger
 \rlap{\hbox to \dimen1{\hfil$#1$\hfil}}   % so center #1
   /                                         % and print /
  \fi}                                         %
\def\cpto{\mathrel {\vcenter {\baselineskip 0pt \kern 0pt
    \hbox{$CP$} \kern 0pt \hbox{$\longrightarrow$} }}}
\def\cptof{\mathrel {\vcenter {\baselineskip 0pt \kern 0pt
    \hbox{$~CP$} \kern 0pt \hbox{$\longleftrightarrow$} }}}

\begin{document}

\baselineskip=15pt

\preprint{}

\title{Unitarity and vacuum stability constraints on the couplings of color octet scalars}

\author{Xiao-Gang He${}^{1,2,3}$\footnote{hexg@phys.ntu.edu.tw}}
\author{Han Phoon$^4$
\footnote{hphoon@iastate.edu}}
\author{Yong Tang${}^{1}$\footnote{ytang@phys.cts.nthu.edu.tw}}
\author{German Valencia$^4$
\footnote{valencia@iastate.edu}}
\affiliation{${}^{1}$Physics Division, National Center for Theoretical Sciences,\\ 
Department of Physics, National Tsing Hua University, Hsinchu, Taiwan}
\affiliation{${}^{2}$INPAC, Department of Physics, Shanghai Jiao Tong University, Shanghai, China}
\affiliation{${}^{3}$CTS, CASTS and Department of Physics, National Taiwan University, Taipei, Taiwan}
\affiliation{${}^{4}$Department of Physics, Iowa State University, Ames, IA 50011, United States}

\date{\today $\vphantom{\bigg|_{\bigg|}^|}$}

\date{\today}

\vskip 1cm
\begin{abstract}

The recent discovery of a 126 GeV boson at the LHC will be followed by a detailed examination of its couplings in order to determine whether this particle is the Higgs boson of the standard model or one of many particles of an extended scalar sector. One such extension with a rich phenomenology consists of a color octet electroweak doublet scalar.  The most general renormalizable scalar potential contains twelve new parameters and it is therefore desirable to constrain them. We present theoretical constraints on these parameters obtained by requiring perturbative unitarity for two-to-two scalar  scattering amplitudes at high energy and vacuum stability.

\end{abstract}

\pacs{PACS numbers: }

\maketitle

\section{Introduction}

The two experiments ATLAS and CMS at the Large Hadron Collider (LHC) have found a new resonant state of mass near 126 GeV \cite{Aad:2012tfa,Chatrchyan:2012ufa}. The properties of this state that have been tested so far match those expected for the standard model, but much work remains to be done to confirm that this is indeed the SM Higgs $h$. Many interesting possibilities for new physics remain open at this point. By measuring the coupling strengths of the new state to the known standard model bosons and fermions the experiments should be able to unravel the nature of the new state.

New physics possibilities remain open even if this new state is the SM Higgs boson because the value of its mass suggests that the vacuum may not be completely stable \cite{Holthausen:2011aa,Xing:2011aa,Degrassi:2012ry,Bezrukov:2012sa,Tang:2013bz}. In the SM, vacuum  stability is closely related to the physical Higgs mass $m_h$ since the quartic self-coupling $\lambda$ in the Higgs potential is related to it by $\lambda = m_h/\sqrt{2}\ v$ ($v=246$~GeV). This quartic coupling must remain positive to guarantee that the Higgs potential is bounded from below.  In the SM the top quark contribution to radiative corrections can drive $\lambda$ negative, induce a false and deep minimum at large field values and destabilize the electroweak vacuum. In the SM with a Higgs mass near 126 GeV,  the turning point is at an energy scale of order $10^{10}$ GeV and there is no immediate conflict with data. However, if we require the vacuum to be stable up to grand unification or even Planck scales, new physics is required to change the running of $\lambda$\cite{Chen:2012faa,EliasMiro:2012ay,Lebedev:2012zw,Rodejohann:2012px,Cheung:2012nb,Kannike:2012pe,Chao:2012mx,Iso:2012jn,Allison:2012qn,Belanger:2012zr,Patel:2012pi,Chao:2012xt,Dev:2013ff,Goudelis:2013uca}. In general this new physics will also affect the phenomenology of Higgs production and decay at the LHC.

In this paper we consider a simple extension of the scalar sector of the SM with new scalars $S$ transforming as $(8,2,1/2)$ under the  SM gauge group $SU(3)_C\times SU(2)_L\times U(1)_Y$. This color octet, electroweak doublet, scalar extension of the SM is motivated by the requirement of minimal flavor violation \cite{Chivukula:1987py,D'Ambrosio:2002ex,Manohar:2006ga}.  The most general renormalizable scalar potential for this model contains twelve new parameters and it is therefore desirable to constrain them. We compute  theoretical constraints on these parameters  by requiring perturbative unitarity for the two-to-two scalar scattering amplitudes at high energy, as well as by requiring vacuum stability in the form of a positive $\lambda$.

In addition to the color octet, the scalar sector of this model contains a Higgs boson from the doublet $H: (1,2,1/2)$ which is responsible for spontaneous electroweak symmetry breaking and has the same tree level couplings as the SM $h$ to other SM particles. The additional colored scalars are assumed to be heavier and the phenomenology related to their possible observation at LHC has been studied before.  The new scalars can induce important loop effects which are substantially different from the SM due to new parameters in the Yukawa and Higgs potential sectors. We will discuss the effects of this extension of the SM on the question of vacuum stability and we will address the phenomenology of the loop level Higgs couplings $hgg$ and $h\gamma\gamma$ in a separate publication \cite{usinprep}.

\section{The model} \label{s:model}

The model we consider is an extension of the scalar sector of the SM to which a color octet, electroweak doublet of scalars is added. The inclusion of the new multiplet $S$ introduces several new, renormalizable, interaction terms to the Lagrangian. Because $S$ has non-trivial $SU(3)_C\times SU(2)_L\times U(1)_Y$ quantum numbers, it will have corresponding gauge interactions. In addition there will be new terms in the Yukawa couplings and in the Higgs potential that are consistent with minimal flavor violation. Following Ref.~\cite{Manohar:2006ga} we write the Yukawa couplings as 
\begin{equation}
\mathcal{L}_Y=-\tilde\eta_U g^{U}_{ij}\bar{u}_{Ri}T^A Q_j S^A - \tilde\eta_D g^{D}_{ij}\bar{d}_{Ri}T^A Q_j S^{\dagger A} + h.c,
\end{equation}
where $Q_i$ are left-handed quark doublets, $S=S^aT^a \;(a=1,...,8)$ with the $SU(3)$ generators normalized as $\Tr (T^aT^b)=\delta^{ab}/2$. The  matrices $g^{U,D}_{ij}$ are the same as those coupling the Higgs to quarks, and $\eta_{U,D}$ are new overall factors that can be complex. In the quark mass eigenstate basis these couplings are given by
\begin{eqnarray}
{\cal L} &=& - {\sqrt{2}\over v} \tilde \eta_{U} \bar U_R T^a \hat M^u U_LS^{a0} + {\sqrt{2}\over v} \tilde \eta_{U} \bar U_R T^a \hat M^u V_{KM} D_L S^{a+}\nonumber\\
&-& {\sqrt{2}\over v} \tilde \eta_{D} \bar D_R T^a \hat M^d D_LS^{a0\dagger} - {\sqrt{2}\over v} \tilde \eta_{D} \bar D_R T^a \hat M^u V^\dagger_{KM} U_L S^{a-} + {\rm ~h.c.},
\end{eqnarray}
where $\hat{M}^{u,d}$ are the diagonal quark mass matrices, $\hat{M}^{u,d} = {\rm diag}(m_{u,d}, m_{c,s}, m_{t,b})$; the quark fields are 
$U_{L,R} = {\rm diag}(u_{L,R}, c_{L,R}, t_{L,R})$ and $D_{L,R} = {\rm
diag}(d_{L,R}, s_{L,R}, b_{L,B})$. The neutral complex field $S^{a0}$ can be further decomposed into a scalar $S^{a0}_R$ and a pseudo-scalar $S^{a0}_I$ as $S^{a0} = (S^{a0}_R + i\ S^{a0}_I)/\sqrt{2}$.
The parameters $\tilde \eta_{U,D}$ are expected to be of order one and are in general complex. We will write them as $\tilde \eta_{U,D} = \eta_{U,D}e^{i\alpha_{u,d}}$ with $\eta_{U,D}$ real, and if there are
non-zero phases $\alpha_{u,d}$ there is $CP$ violation beyond the SM.

The most general renormalizable scalar potential is given in Ref.~\cite{Manohar:2006ga} as\footnote{We use a normalization of $\lambda$ different than Ref.~\cite{Manohar:2006ga} in order to have the conventional relation $\lambda=G_Fm_H^2/\sqrt{2}$.}
\begin{eqnarray}
V&=&\lambda\left(H^{\dagger i}H_i-\frac{v^2}{2}\right)^2+2m_s^2\ {\rm Tr}S^{\dagger i}S_i +\lambda_1\ H^{\dagger i}H_i\  {\rm Tr}S^{\dagger j}S_j +\lambda_2\ H^{\dagger i}H_j\  {\rm Tr}S^{\dagger j}S_i 
\nonumber \\
&+&\left( \lambda_3\ H^{\dagger i}H^{\dagger j}\  {\rm Tr}S_ iS_j +\lambda_4\ e^{i\phi_4}\ H^{\dagger i} {\rm Tr}S^{\dagger j}S_ jS_i +
\lambda_5\ e^{i\phi_5}\ H^{\dagger i} {\rm Tr}S^{\dagger j}S_ iS_j 
+{\rm ~H.c.}\right)\nonumber \\
&+& \lambda_6\  {\rm Tr}S^{\dagger i}S_ iS^{\dagger j} S_j +\ 
 \lambda_7\  {\rm Tr}S^{\dagger i}S_ jS^{\dagger j} S_i +\
  \lambda_8\  {\rm Tr}S^{\dagger i}S_ i\ {\rm Tr}S^{\dagger j} S_j
  \nonumber \\
  &+&  \lambda_9\  {\rm Tr}S^{\dagger i}S_ j\ {\rm Tr}S^{\dagger j} S_i
  +\  \lambda_{10}\  {\rm Tr} S_i
S_ j\ {\rm Tr}S^{\dagger i}S^{\dagger j}+\ 
\lambda_{11}\  {\rm Tr} S_iS_ jS^{\dagger j}S^{\dagger i}.
\label{potential}
\end{eqnarray}
Here $v\sim 246$~GeV is the Higgs
vacuum expectation value (vev) with $\langle H \rangle = v/\sqrt{2}$.
The traces are over the color indices and the $SU(2)$ indices $i,j$ are displayed explicitly. We follow Ref.~\cite{Manohar:2006ga} to pick a real $\lambda_3$ by a suitable definition of $S$ fields and have displayed the two possible phases $\phi_{4,5}$ explicitly, all the $\lambda's$ in Eq.~\ref{potential} are thus real.

Although implementation of minimal flavor violation reduces the parameters in the Yukawa sector, there are still a large number of parameters in the scalar potential $V$. These parameters have physical effects and are constrained from various theoretical and experimental considerations.
In order to have an unbroken  color symmetry, the parameters must be chosen such that $S$ cannot develop a non-zero vev,  but $H$ must have the usual non-zero vev to induce electroweak symmetry breaking. In order for the 
potential to be bounded from below  $\lambda$ must be larger than zero and several of $\lambda_i$s must be positive. Maintaining unitarity of the scattering amplitudes induced by the new parameters also constrains them. 

After symmetry breaking, the non-zero vev of the Higgs gives the physical Higgs scalar $h$  a mass $m^2_H = 2 \lambda v^2$ and it also splits  the octet scalar masses. The resulting tree level mass spectrum for the colored scalars is \cite{Manohar:2006ga}
\begin{eqnarray}
m^2_{S^{\pm}} & = & m^2_S + \lambda_1 \frac{v^2}{4},\nonumber \\
m^2_{S^{0}_R}& = & m^2_S + \left(\lambda_1 + \lambda_2 + 2 \lambda_3 \right) \frac{v^2}{4},\\
m^2_{S^{0}_I}& = & m^2_S + \left(\lambda_1 + \lambda_2 - 2 \lambda_3 \right) \frac{v^2}{4},\nonumber
\end{eqnarray}
The parameters $m_S^2$, and $\lambda_{1,2,3}$ should be chosen such that the above masses remain positive. When not stated explicitly, we shall denote the colored scalar masses collectively by  $M_S$. 

It has been noted that Eq.~\ref{potential} respects the custodial $SU(2)$ symmetry of the SM model if \cite{Manohar:2006ga,Burgess:2009wm,Carpenter:2011yj}
\begin{eqnarray}
2\lambda_3=\lambda_2,\, 2\lambda_6=2\lambda_7=\lambda_{11},\, \lambda_9=\lambda_{10},\, \lambda_4=\lambda_5^\star.
\label{custsym}
\end{eqnarray}
When these conditions are imposed, constraints from the $T$ parameter, from electroweak precision data, are automatically satisfied.

\section{Two-to-two scalar scattering amplitudes and unitarity constraints}\label{s:unitarity}

In this section we consider high energy two-to-two scalar scattering to constrain the strength of the self interactions with the requirement of perturbative unitarity. Although the potential is renormalizable, the tree-level scattering amplitudes approach a constant at high energy that is proportional to the quartic couplings. Perturbative unitarity then constrains their size in a manner entirely analogous to the unitarity bound on the SM Higgs boson mass \cite{Lee:1977eg} and generalizations \cite{Kanemura:1993hm}. We will consider scattering of all the scalar particles that appear in the model at energies much larger than their masses. In this limit, the Higgs and the SM would be Goldstone bosons can be treated as degenerate and massless, and the colored scalars as degenerate with mass $m_S$. The strongest limits on the couplings are obtained by considering scattering of two particle states of definite color and $I=0$.  In this context, $I=0$ is the singlet of the approximate $O(4)$ symmetry between $H,w^\pm,z$ (referred collectively as $h$) and between $S^\pm,S_i,S_r$. The normalization of the  two particle states will then be 
\begin{eqnarray}
|SS\rangle_0 &\equiv& \frac{1}{\sqrt{8}}S^a S^a\nonumber \\
|SS\rangle_{8S} &\equiv& \sqrt{\frac{3}{5}}d_{abc}S^a S^b\nonumber \\
|SS\rangle_{8A} &\equiv& \frac{1}{\sqrt{3}}f_{abc}S^a S^b,
\end{eqnarray}
for the color singlet, the symmetric octet and the  antisymmetric octet respectively. 
The $I=0$ two particle state will be 
\begin{eqnarray}
|hh\rangle_{I=0} &\equiv& \frac{1}{\sqrt{8}}(w^+w^-+w^-w^++zz+hh) \nonumber \\
|SS\rangle_{I=0} &\equiv& \frac{1}{\sqrt{8}}(S^{a+}S^{b-}+S^{a-}S^{b+}+S_i^{a}S_i^{b}+S_r^{a}S_r^{b}) \nonumber \\
|hS\rangle_{I=0} &\equiv& \frac{1}{\sqrt{8}}(w^+S^{c-}+w^-S^{c+}+S_i^{c}z+S_r^{c}h)
\end{eqnarray}
for two color singlet scalars, two color octet scalars or one color singlet and one color octet respectively.

The best unitarity bound on the Higgs boson mass arises from considering the high energy limit of $hh$ scattering in the $I=0$ and $J=0$ partial wave \cite{Lee:1977eg}. In the same manner we begin by considering the isospin zero, $J=0$ scattering of $hh\to SS$ which is also a color singlet. We find a partial wave amplitude that depends only on the parameters $\lambda_{1,2,3}$ (using the notation $a_{IJ}^c$) given by
\begin{eqnarray}
a_{00}^0&=& a(|hh\rangle_{I=0} \to |SS\rangle_{I=0}^{0})_{J=0} \nonumber \\
&=&\frac{1}{8\sqrt{2}\pi}(2\lambda_1+\lambda_2)+\frac{3}{8\sqrt{2}\pi}\frac{v^2}{s-m_H^2}\lambda(2\lambda_1+\lambda_2) \nonumber \\
&+& \frac{1}{16\sqrt{2}\pi}\frac{v^2}{s\beta_z\beta_s}\left(\lambda_1^2+\lambda_1\lambda_2+2\lambda_2^2\right)\log\left(\frac{s(1-\beta_z\beta_s)-2m_z^2}{s(1+\beta_z\beta_s)-2m_z^2}\right) \cdots, 
\label{vvss}
\end{eqnarray}
where we have defined $\beta_i\equiv \sqrt{1-4m_i^2/s}$ and $\cdots$ stands for terms proportional to mass splittings. In the high energy limit only the first term (due to a contact interaction) remains, and requiring that ${\rm Re}(a_{00}^0)<1/2$, results on the perturbative unitarity constraint
\begin{equation} 
\left| 2\lambda_1+\lambda_2\right| \lsim 18
\label{l12bound} 
\end{equation}
We note that this constraint does not use the custodial symmetry relations Eq.~\ref{custsym}, as $\lambda_3$ only appears in the terms with the logarithm which vanish in the high energy limit. 

Without custodial symmetry, we can consider all the separate $hh\to SS$, $J=0$, partial wave amplitudes in the color singlet channel to obtain separate bounds for the couplings $\lambda_{1,2,3}$ and this is shown in Table~\ref{t:hhss} in the appendix.  When constraining one coupling at a time, none of the results in Eq.~\ref{morehhss} improves on the condition Eq.~\ref{l12bound}, but we do obtain new information in the form 
\begin{equation}
\lambda_3 \lsim 8.9.
\label{l3bound}
\end{equation}

Next we consider the scattering  $hS \to SS$ which is proportional to $\lambda_{4,5}$ in the high energy limit. Our best bound arises from  the isospin zero, zeroth partial wave, symmetric color octet channel where we find in the custodial symmetry limit
\begin{eqnarray}
a_{00}^{8S}&=& a(|hS\rangle_{I=0} \to |SS\rangle_{I=0}^{8S})_{J=0} \nonumber \\
&=&\frac{1}{32\pi }\sqrt{15}\  \lambda_4\cos\phi_4\ \left(1+\frac{v^2}{8(s-m_s^2)}(\lambda_1+10\lambda_3)\right. \nonumber \\
&-& \left. \frac{v^2}{12\beta_s(s-m_s^2)} (3\lambda_1+12\lambda_3) \log\left(\frac{1+\beta_s}{1-\beta_s}\right)\right)+\cdots
\label{vsss}
\end{eqnarray}
where once again the $\cdots$ stands for terms involving mass splittings. In the high energy limit this results in the bound
\begin{eqnarray}
\left| \lambda_4\cos\phi_4+\lambda_5\cos\phi_5\right| \lsim 26
\label{l45bound}
\end{eqnarray}
which we have re-written in terms of both $\lambda_{4,5}$ as it appears without assuming custodial symmetry.

The remaining coupling constants, $\lambda_{4-11}$, affect only the colored scalars self-interactions at tree-level and we constrain them by looking at $SS$ scattering. In the custodial symmetry limit, a  first bound is obtained from the $I=0$ color singlet channel
\begin{eqnarray}
a^0_{00} &=& a(|SS\rangle_{|I=0>}^0 \to |SS\rangle_{|I=0>}^0)_{J=0} \nonumber \\
&=&\frac{1}{32 \pi }\left( 17\lambda_8+13\lambda_{9}+13\lambda_{11} 
\right) +\frac{1}{4\pi}\frac{v^2}{s-m_H^2}(\lambda_1+\lambda_3)^2\nonumber \\
&-&\frac{1}{64 \pi }\frac{v^2}{(s-4m_s^2)}\left((\lambda_1^2+4\lambda_3^2+2\lambda_1\lambda_3)\log\left(\frac{s-4m_s^2+m_H^2}{m_H^2}\right)\right)\nonumber \\
&-&\frac{3}{128 \pi } \frac{v^2}{(s-4m_s^2)}\left(\lambda_4^2(4+\cos(2\phi_4))\log\left(\frac{s-3m_s^2}{m_s^2}\right)\right)\nonumber \\
&-&\frac{1}{32 \pi } \frac{v^2}{(s-4m_s^2)}\lambda_3^2\left(
\log\left(\frac{s-4m_s^2+m_z^2}{m_z^2}\right)+2\log\left(\frac{s-3m_s^2}{m_s^2}\right)\right) + \cdots
\label{ssss00}
\end{eqnarray}
The tightest constraint in the custodial symmetry limit  arises from Eq.~\ref{ssss00} and is given by,
\begin{eqnarray}
\left| \frac{1}{32 \pi }\left( 17\lambda_8+13\lambda_{9}+13\lambda_{11} 
\right) \right| &<& \frac{1}{2}.
\label{octetb1}
\end{eqnarray}
Using this result to bound one coupling at a time  gives
\begin{eqnarray}
\left| \lambda_8\right| \lsim 3, && \left| \lambda_{9,11}\right| \lsim 3.9
\end{eqnarray}

A second independent constraint is obtained by considering scattering in the symmetric color octet channel
\begin{eqnarray}
a^{8S}_{00}&=&a(|SS\rangle_{|I=0>}^0 \to |SS\rangle_{|I=0>}^{8S})_{J=0} \nonumber \\
&=&\frac{1}{64\pi}(2\lambda_8+10\lambda_9+7\lambda_{11})+\frac{15}{128\pi}\frac{v^2}{(s-m_s^2)}\lambda_4^2\cos^2(\phi_4)
\nonumber \\
&+&\frac{9}{256\pi}\frac{v^2}{(s-4m_s^2)}\lambda_4^2\cos(2\phi_4)\log\left(\frac{s-3m_s^2}{m_s^2}\right)\nonumber \\
&-& \frac{1}{64\pi} \frac{v^2}{(s-4m_s^2)}(\lambda_1^2+4\lambda_3^2+2\lambda_1\lambda_3)\log\left(\frac{s-4m_s^2+m_H^2}{m_H^2}\right)\nonumber \\
&-&\frac{1}{32\pi} \frac{v^2}{(s-4m_s^2)}\lambda_3^2\left(
\log\left(\frac{s-4m_s^2+m_z^2}{m_z^2}\right)+2\log\left(\frac{s-3m_s^2}{m_s^2}\right)\right)+ \cdots
\label{ssss08}
\end{eqnarray}
from which it follows that
\begin{eqnarray}
\left| \frac{1}{64 \pi }\left( 2\lambda_8+10\lambda_{9}+7\lambda_{11} 
\right) \right| &<& \frac{1}{2}.
\label{octetb2}
\end{eqnarray}

The antisymmetric color octet channel vanishes, as expected, for the symmetric $I=0$ state. It produces additional constraints  if we consider amplitudes that are antisymmetric in $O(4)$ indices or by considering the scattering of specific channels away from the custodial symmetry limit. These results are shown in the Appendix. 

Finally, we can also consider $t\bar{t} \to t\bar{t} $ scattering at high energy to constrain $\eta_U$. 
Consider the zeroth partial wave for elastic scattering of a same helicity $t\bar{t}$ state in the color singlet and color octet channels. The color singlet channel is responsible for the well known SM result $a_0(t\bar{t} \to t\bar{t}) \to \frac{3}{8\pi}\frac{m_t^2}{v^2}$. In this model it does not provide any constraint as it receives contributions from $t$-channel exchange of  $S_r$ and $S_i$ that cancel each other out. The color octet channel, on the other hand gives us at high energy 
\begin{eqnarray}
a_0^8(t\bar{t} \to t\bar{t}) = |\eta_U|^2 \frac{1}{16\pi}\frac{m_t^2}{v^2}
\end{eqnarray}
which results in the rather weak bound
\begin{eqnarray}
|\eta_U| \lsim  \ 7.2
\end{eqnarray}

\section{The running of $\lambda$}\label{s:stability}

We now study the effect of the color octet scalars on the running of $\lambda$ and consequently on vacuum stability. As mentioned before there are many parameters in the potential and this makes a general analysis complicated. At the leading one loop level however, only $\lambda_{1,2,3}$  affect the running of $\lambda$ directly.  In this section we will carry out our analysis in terms of the four new free parameters $m^2_S$ and $\lambda_{1,2,3}$. 
  
The running of these parameters is determined by renormalization group equations (RGE). We find that at one loop level, the $\beta$ function governing the running of the quartic coupling $\lambda$  is given by
\begin{align}\label{eq:betalambda}
\beta_{\lambda} =& \frac{1}{16\pi^2}\Big{[}24 \lambda^2 + 4 \lambda_1^2 +4\lambda_1\lambda_2+ 2 \lambda_2^2 + 8\lambda_3^2 -(3g'^2+9g^2-12 y_t^2)\lambda  \nonumber \\
& {}- 6 y_t^4 + \frac{3}{8}g'^4 + \frac{3}{4} g'^2 g^2 +\frac{9}{8} g^4 \Big{]} ,
\end{align}
where $g_s$, $g$ and $g'$ are associated with $SU(3)_C$, $SU(2)_L$ and $U(1)_Y$, respectively and  $y_t = g^U_{33}= \sqrt{2}m_t/v$ is the top quark Yukawa coupling. 

It is easy to see that octets always contribute positively to $\beta_{\lambda}$ for non vanishing $\lambda_i$ since 
\[
 4 \lambda_1^2 +4\lambda_1\lambda_2+ 2 \lambda_2^2 + 8\lambda_3^2 = \left(2\lambda_1+\lambda_2\right)^2 + \lambda_2^2 + 8\lambda_3^2 \geq 0.
\]
Note that the above RGE is valid when $\mu\geq M_S$. If $\mu \ll M_S$, the decoupling theorem tells us that the effects of the octet can be neglected and terms with $\lambda_i$ do not contribute, leading effectively to the SM. 

The $\beta$ functions for $\lambda_{1,2,3}$ are given by (we include contributions to these from $\lambda_{4-11}$  in the custodial symmetry limit but give general expressions in the Appendix)
\begin{eqnarray}\label{eq:betalambda1}
\beta_{\lambda_1} &=&\frac{1}{16\pi^2}\Big{[} 2 \lambda_1^2 +\lambda_2^2 + 4\lambda_3^2 + 4 \lambda(3 \lambda_1+\lambda_2) - (\frac{3}{2}g^{'2}+ \frac{9}{2}g^{2} - 6 y_t^2 )\lambda_1 \nonumber \\
&&{} - (\frac{3}{2}g^{'2}+\frac{9}{2} g^2 + 18 g_s^{2} -  \eta_U^2 y_t^2 ) \lambda_1 - 4 \eta_U^2 y_t^4 + \frac{3}{2} g^{'4} - 3  g^{'2}g^2 + \frac{9}{2} g^4 \nonumber \\
&& +\frac{10}{3}\lambda_4^2\cos^2\phi_4+\lambda_1\left(13\lambda_{11}+17\lambda_8+13\lambda_9\right)+\frac{2}{3}\lambda_2\left(8\lambda_{11}+12\lambda_8+3\lambda_9\right) \Big{]},
\end{eqnarray}
\begin{eqnarray}\label{eq:betalambda2}
\beta_{\lambda_2} &=&\frac{1}{16\pi^2}\Big{[} 2 \lambda_2^2 + 4\lambda_1\lambda_2+ 16\lambda_3^2 + 4 \lambda \lambda_2 - (\frac{3}{2}g^{'2}+ \frac{9}{2}g^{2} - 6 y_t^2)\lambda_2 \nonumber \\
&&{} - (\frac{3}{2}g^{'2}+\frac{9}{2} g^2 + 18 g_s^{2} -  \eta_U^2 y_t^2) \lambda_2 - 4 \eta_U^2 y_t^4 + 6 g^{'2}g^2 \nonumber \\
&& +\frac{25}{6}\lambda_4^2\cos^2\phi_4+\frac{1}{3}\lambda_2\left(7\lambda_{11}+3\lambda_8+27\lambda_9\right)
\Big{]},
\end{eqnarray}
and
\begin{eqnarray}\label{eq:betalambda3}
\beta_{\lambda_3} &=&\frac{1}{16\pi^2}\Big{[} 2 \lambda_3(2\lambda + 2\lambda_1 + 3\lambda_2) - (\frac{3}{2}g^{'2}+ \frac{9}{2}g^{2} - 6 y_t^2)\lambda_3 \nonumber \\
&&{} - (\frac{3}{2}g^{'2}+\frac{9}{2} g^2 + 18 g_s^{2} -  \eta_U^2 y_t^2) \lambda_3 - 2 \eta_U^2 y_t^4 \nonumber \\
&& + \frac{25}{12}\lambda_4^2\cos^2\phi_4+\frac{1}{3}\lambda_3\left(7\lambda_{11}+3\lambda_8+27\lambda_9\right)
\Big{]}.
\label{beta-lam123}
\end{eqnarray}

It should be pointed out that in the case when $\lambda_i$ vanishes  at tree level, the above RGEs show that gauge and Yukawa interactions can still induce $\lambda_i$ at one loop. Inclusion of $\lambda_i$ is just a necessity of renormalization. $\lambda_3$ is different from $\lambda_{1,2}$ in the sense that if $\eta_U=0$, $\lambda_3$ would not get renormalized at one loop for a vanishing initial input. 

When the energy scale $\mu\geq M_S$, the $\beta$ functions of the gauge couplings also need to be modified as,
\begin{align}
\beta_{g'}  &=\frac{1}{16\pi^2}\left(+\frac{41}{6}+\frac{1}{3}\times\frac{1}{2}\times 8 \right),\\
\beta_{g }  &=\frac{1}{16\pi^2}\left(-\frac{19}{6}+\frac{1}{3}\times\frac{1}{2}\times 8 \right), \\
\beta_{g_s} &=\frac{1}{16\pi^2}\left(-7+\frac{1}{3}\times 3 \right). 
\end{align}
where the first term in each parenthesis is the SM contribution. For the new contributions there is a  factor ${1}/{3}$ for scalars, factors ${1}/{2}$ and $3$  for fundamental and  adjoint representations respectively,  and a factor $8$ for the number of  doublets (one for each color). Finally, the Yukawa coupling has a beta function given by
\begin{align}
\beta_{y_t}=\frac{1}{16\pi^2}y_t\Big{[}\frac{9}{2} y_t^2 + \frac{4}{3}\times\frac{3}{2}\eta_U^2 y_t^2-\frac{9}{4} g^2-\frac{17
   }{12}g'^2-8 g_s^2\Big{]},
\end{align}
\begin{equation}
\beta_{\eta_U}=\frac{1}{16\pi^2}\eta_U\Big{[} \frac{1}{2}\eta_U^2y_t^2-3y_t^2-\frac{9}{2}g^2_s\Big{]}.
\end{equation}

Now that we have all the $\beta$ functions we can study the energy scale dependence of the couplings. For our numerical analysis, we will assume a custodial $SU(2)$ symmetric initial condition, Eq.~\ref{custsym}.   For this study the effect of $\lambda_3$ on the SM Higgs sector is not fundamentally different from that of $\lambda_2$. The special choice $2\lambda_3=\lambda_2$ represents one typical parameter subspace. Also we shall ignore the contributions from $\lambda_{4-11}$ in numerical analysis.  The reason is that these terms affect the running of $\lambda$ only through $\lambda_{1,2,3}$ as secondary effects. It is possible to adjust the input values of $\lambda_{1,2,3}$ to incorporate $\lambda_{4-11}$ contributions.

The initial values of $\lambda_{1,2,3}$ (at a scale of 1~TeV) should not be much larger than $\mathcal{O}(1)$ and should  lie in the ranges of unitarity bounds discussed in Sec.~{\ref{s:unitarity}. If $\lambda_{1,2,3}$ are too large, the perturbative framework is not valid anymore \cite{Bulava:2012rb}, and a study of the resulting non-perturbative effects is beyond the scope of our discussion. If the theory is to remain perturbative up to a given scale, the running of $\lambda_i$ should not reach the unitarity bound and certainly should not reach the Landau pole before that given scale. 

In the following we discuss the running of $\lambda$. We show in Fig.~(\ref{fig:variouslambda}) how the scale of $M_S$ can affect the running of $\lambda$. The solid line shows the original $\lambda$  running behavior in the SM and it is evident that $\lambda$ turns negative at a scale near $10^{10}$GeV {}\footnote{The precise scale depends on the mass of top quark $m_t$ and strong coupling $\alpha_s$, a recent discussion on this can be found in Ref.~\cite{Alekhin:2012py,Masina:2012tz}.}. Now suppose that $M_S\simeq 1\TeV$, the effects of the color octet come into play when $\mu \geq M_S$, as shown in the dotted line. For $\eta_U=0$, we have chosen $\lambda_i = 0$ at $M_S$ as illustration because this choice produces the minimal contributions to $\lambda$. It can still make $\lambda$ positive, or the electroweak vacuum stable, up to the Planck scale. If $M_S$ is very large, the color octet contributions come in too late and are not large enough to stabilize the vacuum, as shown in the dashed line. However, this could be amended very easily by non-vanishing $\lambda_i$ as shown in the dot-dashed line which also shows a threshold effect.

It is interesting to note that even with $\lambda_i = 0$ for $\eta_U=0$, the color octet still modifies $\beta_\lambda$ through   $\beta_{g_i}$ contributions as can be seen in the right panel of Fig.~(\ref{fig:variouslambda}). Color octet effects will make the strong gauge couplings decrease slower than in the SM, and this serves as an additional positive contribution to $\beta_\lambda$. Also, when the gauge couplings are larger, $\lambda_{1,2}$ would run away from zero faster because of the corresponding terms in  Eq.~(\ref{eq:betalambda1}) and Eq.~(\ref{eq:betalambda2}). If the initial values of $\lambda_i$ are not zero, the $\lambda_i^2$ terms in $\beta_\lambda$ give additional positive contributions and may be dominant in the running behavior of $\lambda$.

A non-vanishing $\eta_U$ has a negative contribution to the beta function of $\lambda$. In addition the positive contribution $\frac{4}{3}\times\frac{3}{2}\eta_U^2 y_t^2$ in $\beta_{y_t}$ makes $y_t$  increase faster. This causes the top quark to decrease $\lambda$ more effectively as the energy goes up. We show the running behavior of $\lambda$ for $\eta_U=1$ by the dot-dashed line in Fig.~(\ref{fig:variouslambda}). As can be seen from the figure, if $M_S$ is around the TeV scale and $\lambda_i=0$, the effects of enlarging $\eta$ would decrease the instability scale relative to that in the SM. The end result emerges from competition between non-vanishing $\eta_U$ and $\lambda_i$, and requiring $\lambda_{M_P}>0$ (at the Planck mass)  can constrain their ranges. We illustrate this with an example in Fig.~{\ref{fig:competiton}} for some initial values of $\lambda_1$ and $\eta_U$. 

\begin{figure}[thb]
\includegraphics[width=0.48\textwidth,height=2.85in]{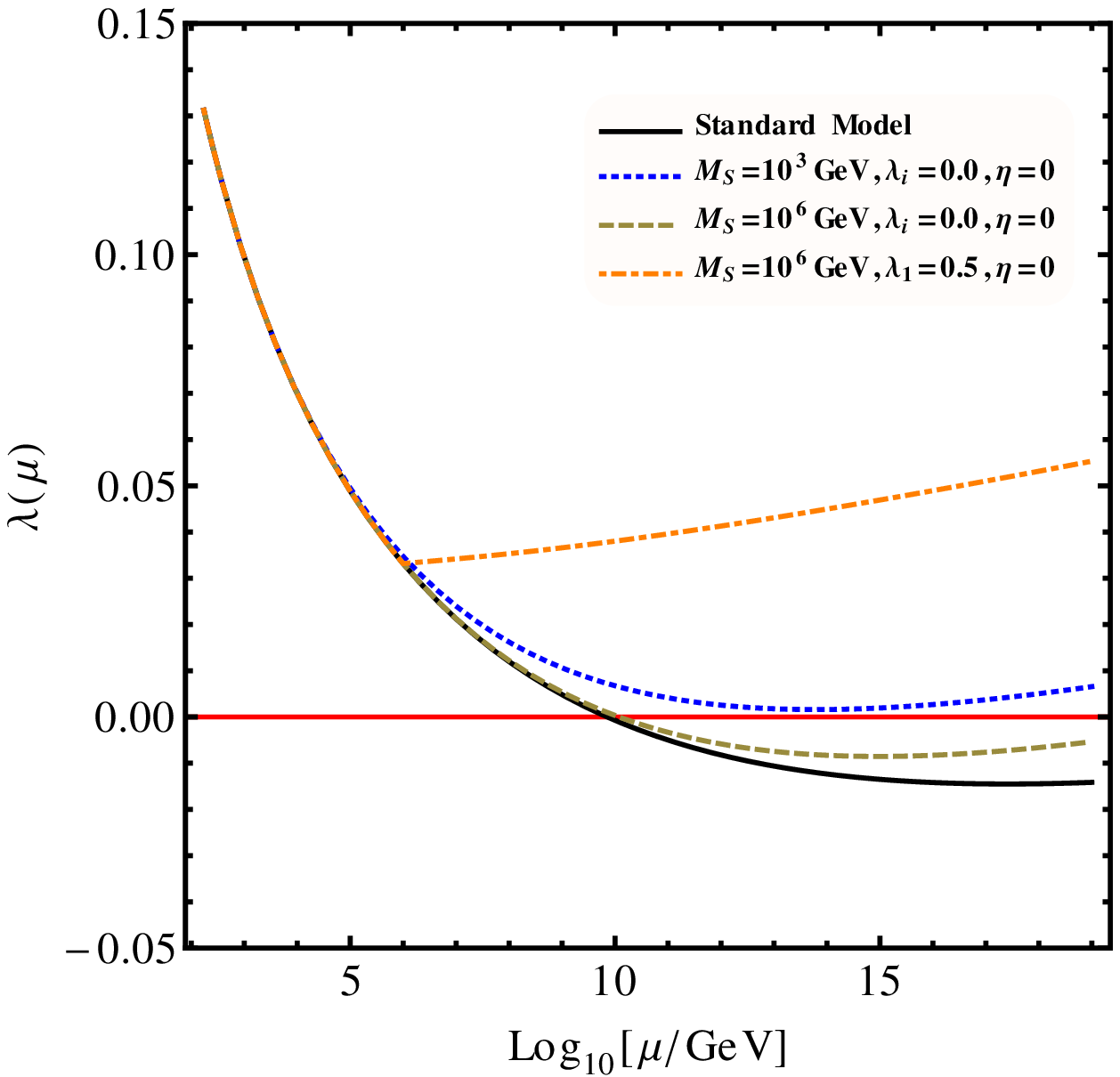}
\includegraphics[width=0.48\textwidth,height=2.85in]{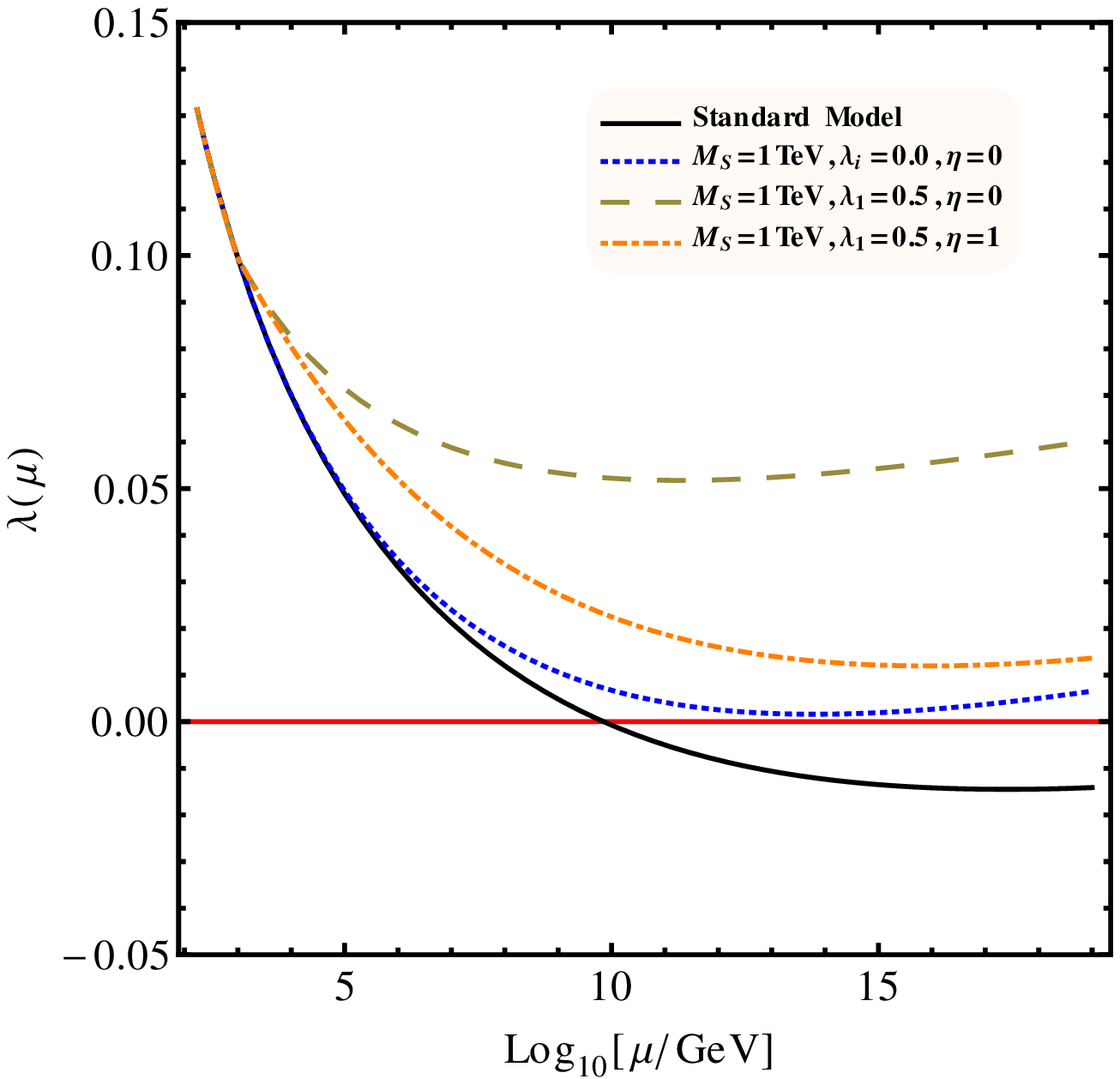}
\caption{
\label{fig:variouslambda}The running of the higgs quartic coupling in the standard model (solid line) and in the octet extension with various values of $\lambda_i$ and $\eta_U$. The figure on the left  shows the effect of the scale $M_S$ and the figure on the right shows the effect of sample values of $\lambda_i$ and $\eta \equiv \eta_U$. }
\end{figure}

\begin{figure}[thb]
\includegraphics[width=0.48\textwidth,height=2.85in]{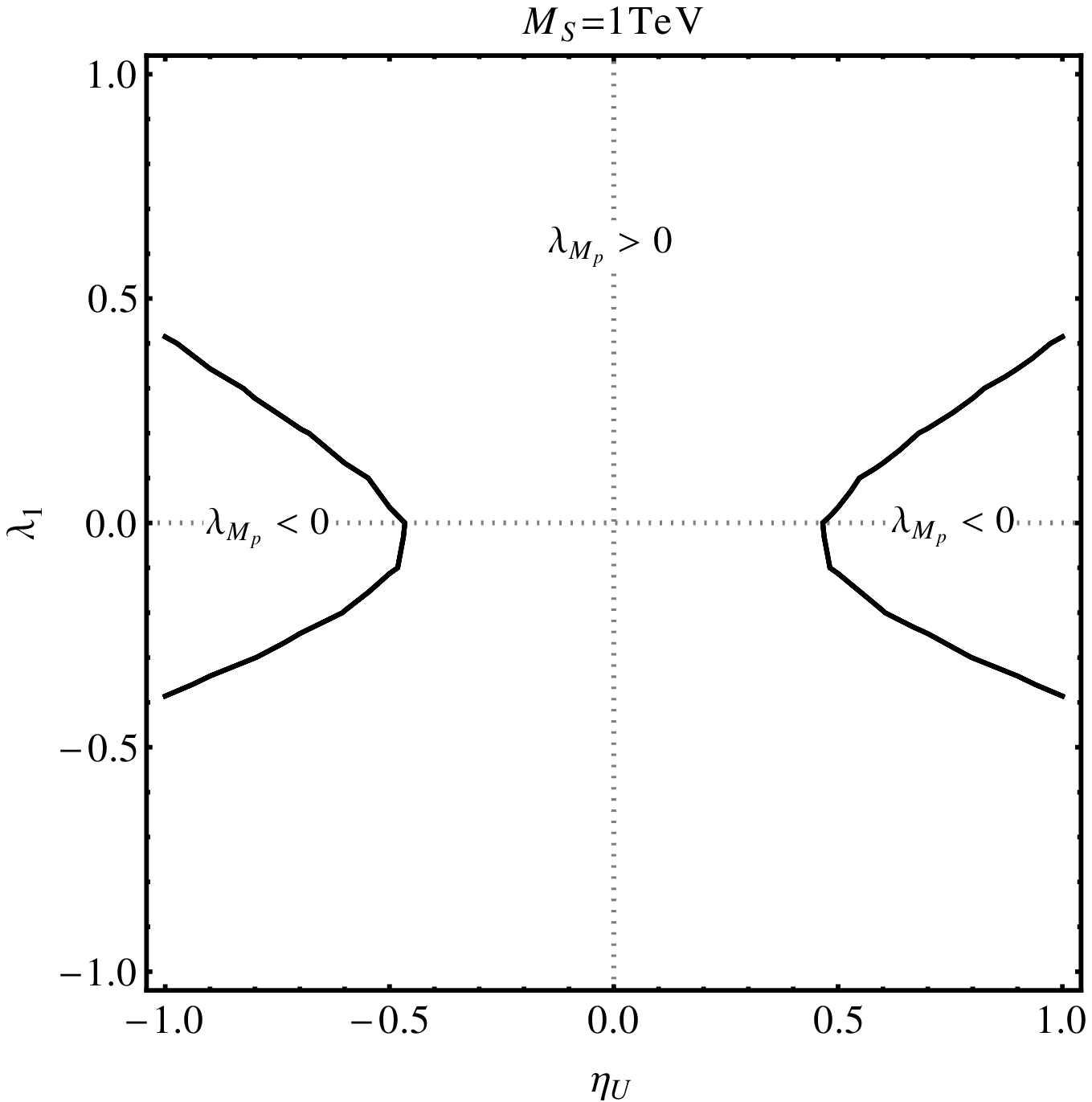}
\includegraphics[width=0.48\textwidth,height=2.85in]{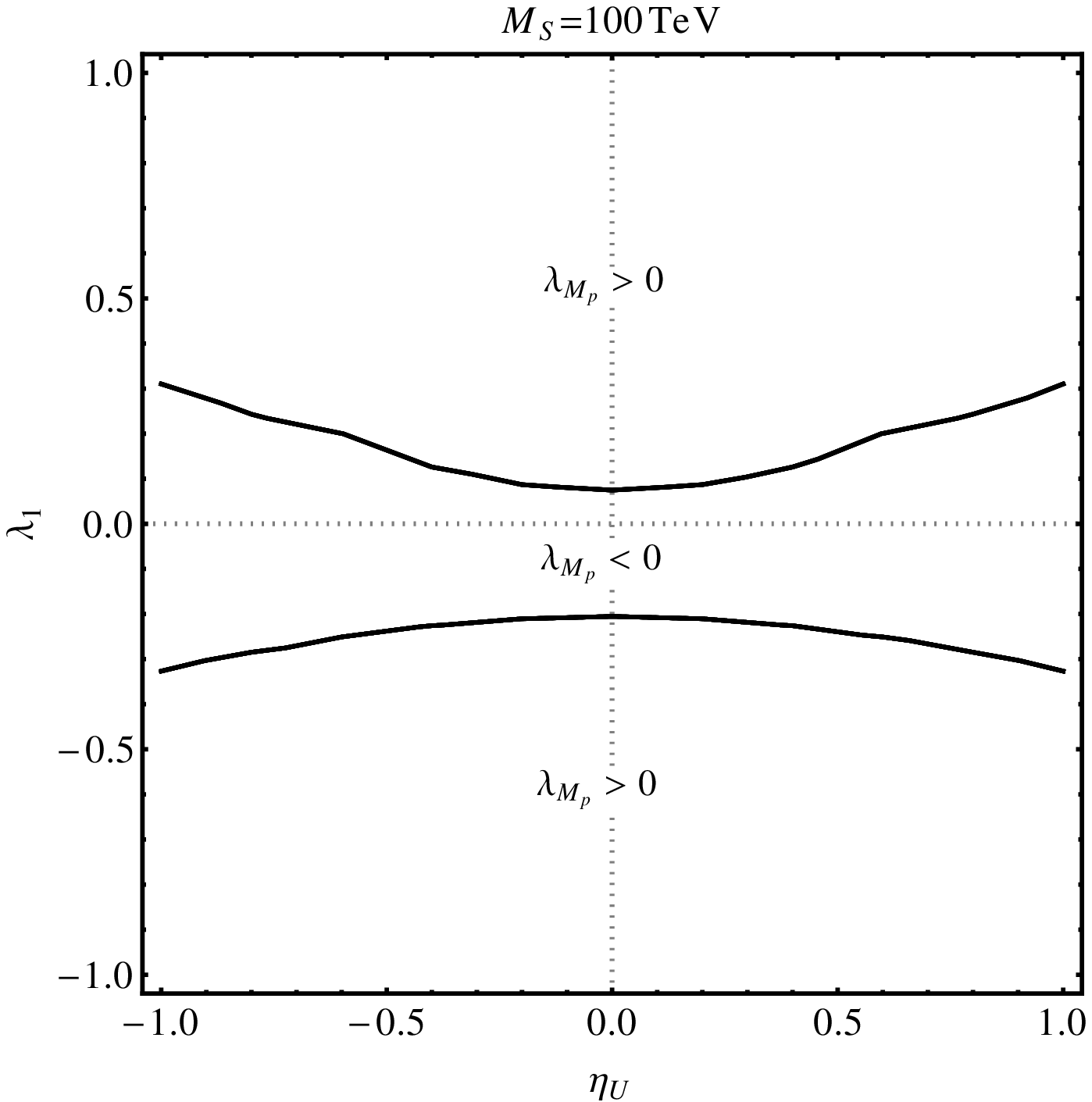}
\caption{
\label{fig:competiton}Sign of $\lambda_{M_p}$ ($\lambda$ at the Planck scale) as contours on $\eta_U$ and $\lambda_1$ assuming vanishing $\lambda_{2,3}$ for two values of $M_S$. Similar constraints exist for $\eta_U$ and $\lambda_2$.}
\end{figure}

\section{Renormalization group improved unitarity bounds}\label{s:renimp}

We can improve the unitarity constraints on the scalar self couplings obtained in Section~\ref{s:unitarity} in some cases by considering their renormalization group evolution along the lines described in Ref.~\cite{Chanowitz:1978uj,Marciano:1989ns} for the Higgs boson mass. In order to do this we need a complete set of one-loop beta functions which we provide in  Appendix~\ref{fullbetas}. 
If we consider only one non-zero coupling at a time, that is we put all other $\lambda_j=0\ (j\neq i)$ in $\beta_{\lambda_i}$, then except for $\lambda_4$, the $\lambda_i$ increase with energy and the corresponding unitarity bounds will be stronger. 
The solution of the complete set of coupled RGE is beyond the scope of this paper, but we 
illustrate the constraints that result in two restricted cases.

We first look at the sector of the potential responsible for the high energy $hh\to SS$ scattering at tree level,  which in the limit of custodial symmetry is governed by the couplings $\lambda_{1,2}$. The relevant RGE are 
\begin{eqnarray}
\frac{d\lambda_1}{d\ln \mu}&=& \frac{1}{8\pi^2}\left( \lambda_1^2 +\lambda_2^2 \right) \nonumber\\
\frac{d\lambda_2}{d\ln \mu}&=& \frac{1}{8\pi^2}\lambda_2 \left( 2\lambda_1 +3\lambda_2 \right).
\end{eqnarray}
These two coupled equations can be readily solved in terms of 
\begin{eqnarray}
a_\pm = \lambda_1+\left(\frac{3}{2}\pm\frac{\sqrt{13}}{2}\right)\lambda_2,
\label{decoup12}
\end{eqnarray}
for which
\begin{eqnarray}
a_\pm(\mu) =\frac{a_\pm(\mu_0)}{1-\frac{1}{8\pi^2}a_\pm(\mu_0)\ln(\frac{\mu}{\mu_0})}.
\label{decoup12run}
\end{eqnarray}
In Fig.~\ref{figl1-l2} we show the region in the $\lambda_{1}-\lambda_2$ (at 1~TeV) plane that satisfies the unitarity constraint, Eq.~\ref{l12bound} up to different energy scales. The red region corresponds to Eq.~\ref{l12bound} for the couplings at 1~TeV indicating how there is no constraint along the direction $\lambda_2=-2\lambda_1$. In the green shaded region we require that the unitarity constraint be satisfied up to 100~TeV and in the blue shaded region up to $10^{10}$~GeV. We see that as we require the theory to remain perturbative to higher scales, the allowed region for positive $\lambda_{1,2}$ shrinks as expected. On the other hand this does not happen for a region where one or both $\lambda_{1,2}$ are allowed to be negative.  In Fig.~\ref{figl1-l2} we also show the conditions  $a_\pm=0$ with the dashed lines. 
 
\begin{figure}[thb]
\includegraphics[width=0.4\textwidth]{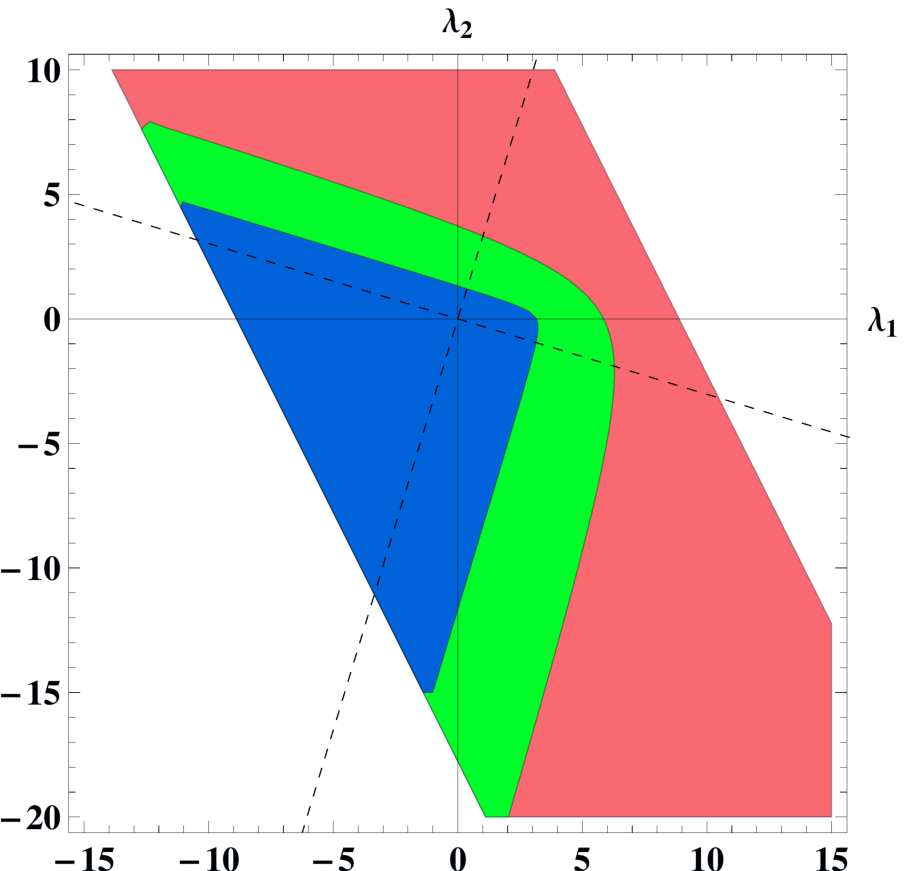}
\includegraphics[width=0.4\textwidth]{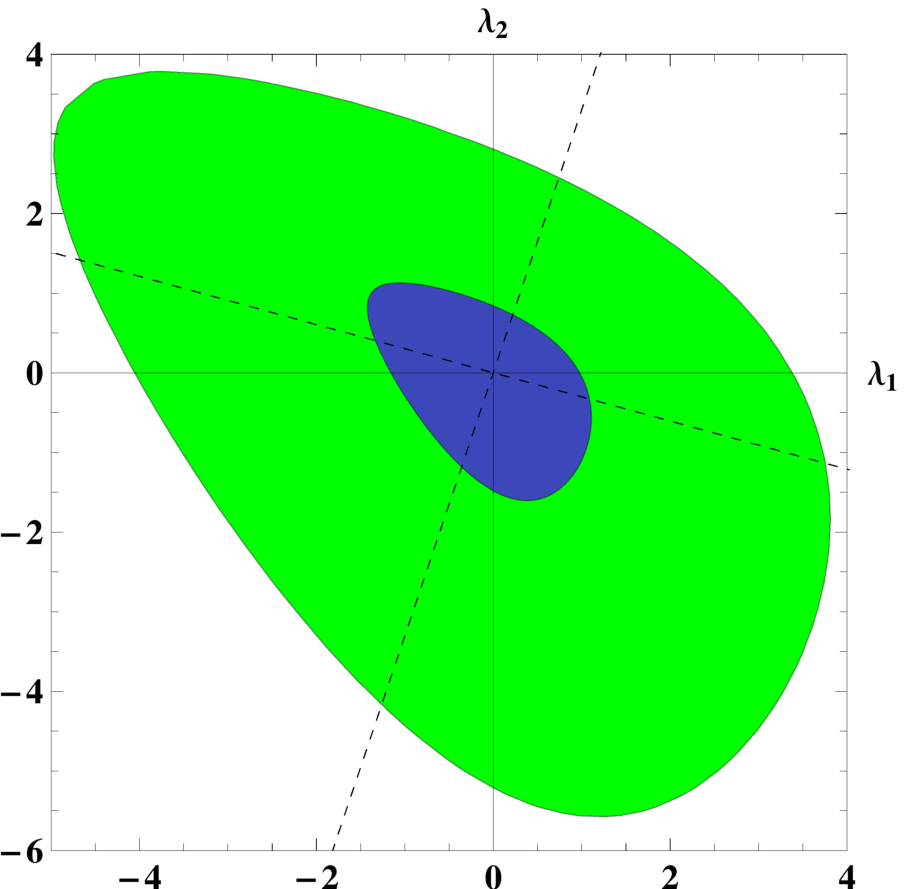}
\caption{
\label{figl1-l2}The left plot shows the region in the $\lambda_1-\lambda_2$ (at 1~TeV) plane that satisfies the unitarity constraint, Eq.~\ref{l12bound} at 1~TeV in red, up to 100~TeV in green and up to $10^{10}$~GeV in blue. The dashed lines show the conditions $a_\pm=0$. In the right plot, this region is further reduced by requiring unitarity up to 100~TeV in green and up to $10^{10}$~GeV in blue for the $\lambda-\lambda_{1,2}$ coupled system.}
\end{figure}

Next we recall that $\lambda_{1,2}$ contribute to the running of $\lambda$ as in Eq.~\ref{eq:betalambda}. We can solve numerically the coupled equations for $\lambda$ and $\lambda_{1,2}$ by ignoring the other couplings and taking the custodial limit. If we then  require that $\lambda$ also satisfy its unitarity constraint \cite{Marciano:1989ns} 
$5\lambda/(16\pi) \leq 1/2$, we further restrict the allowed parameter space. The area in the $\lambda_1-\lambda_2$ plane that is allowed in this case is shown in the right side of Fig.~\ref{figl1-l2}. 

In the above discussion we have neglected the gauge coupling contributions to the RGE. The effects of the gauge couplings, dominated by the strong coupling, tend to slow down the raising of $\lambda_{1,2}$ which in turn slows down the growing rate for $\lambda$. Therefore one expects that inclusion of the gauge couplings will delay the reach of the unitarity bounds. In Fig.~{\ref{fig:lambdairunning}} we use two sets of typical initial values from the blue region in Fig.~\ref{figl1-l2} as illustration, to show the running behavior of $\lambda$ and $\lambda_{1,2}$. We see that indeed $\lambda$ and $\lambda_{1,2}$ are below the unitarity bounds all the way up to a scale higher than $10^{10}$~GeV.

\begin{figure}[tb]
\includegraphics[width=0.48\textwidth,height=2.8in]{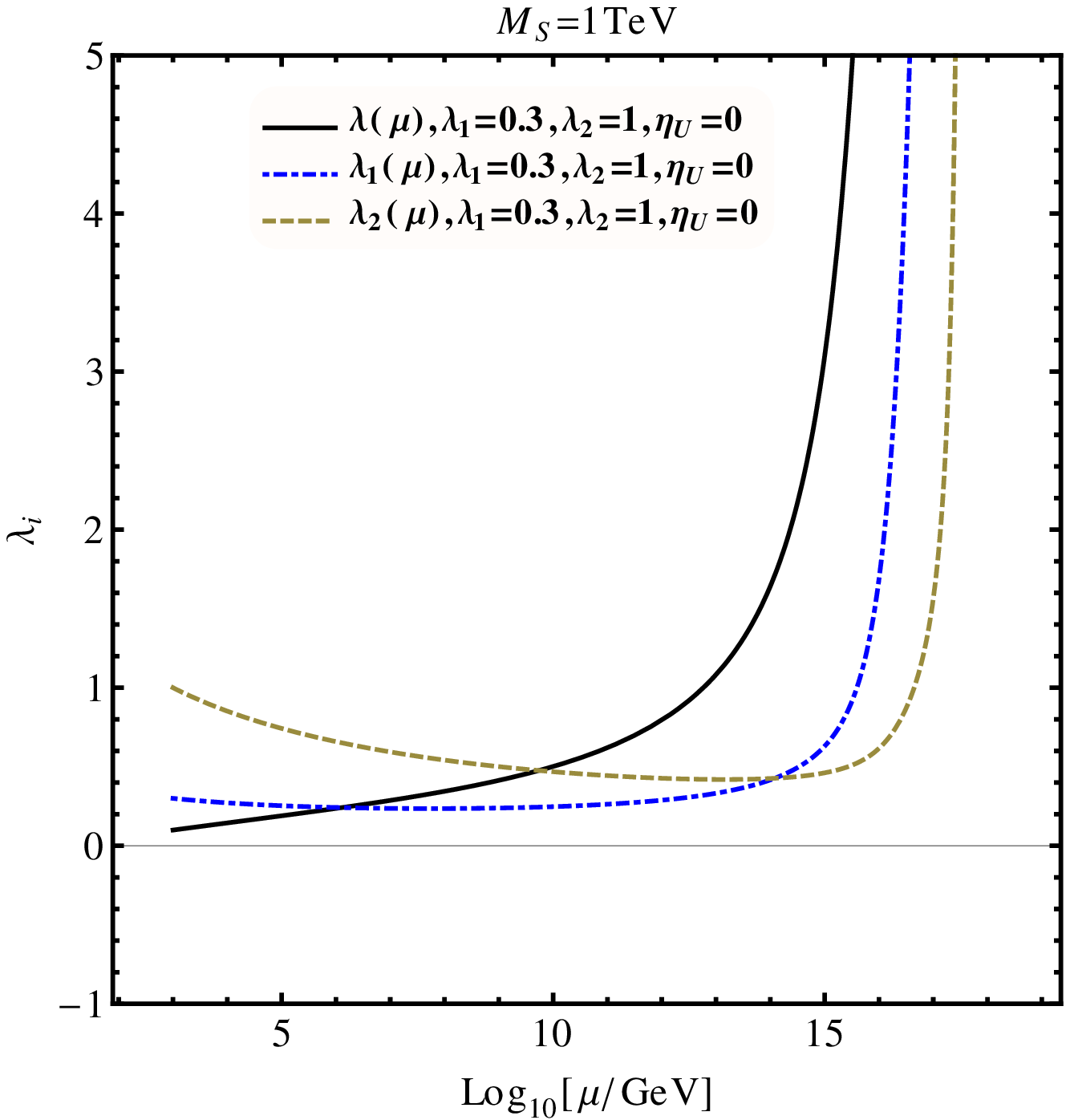}
\includegraphics[width=0.48\textwidth,height=2.8in]{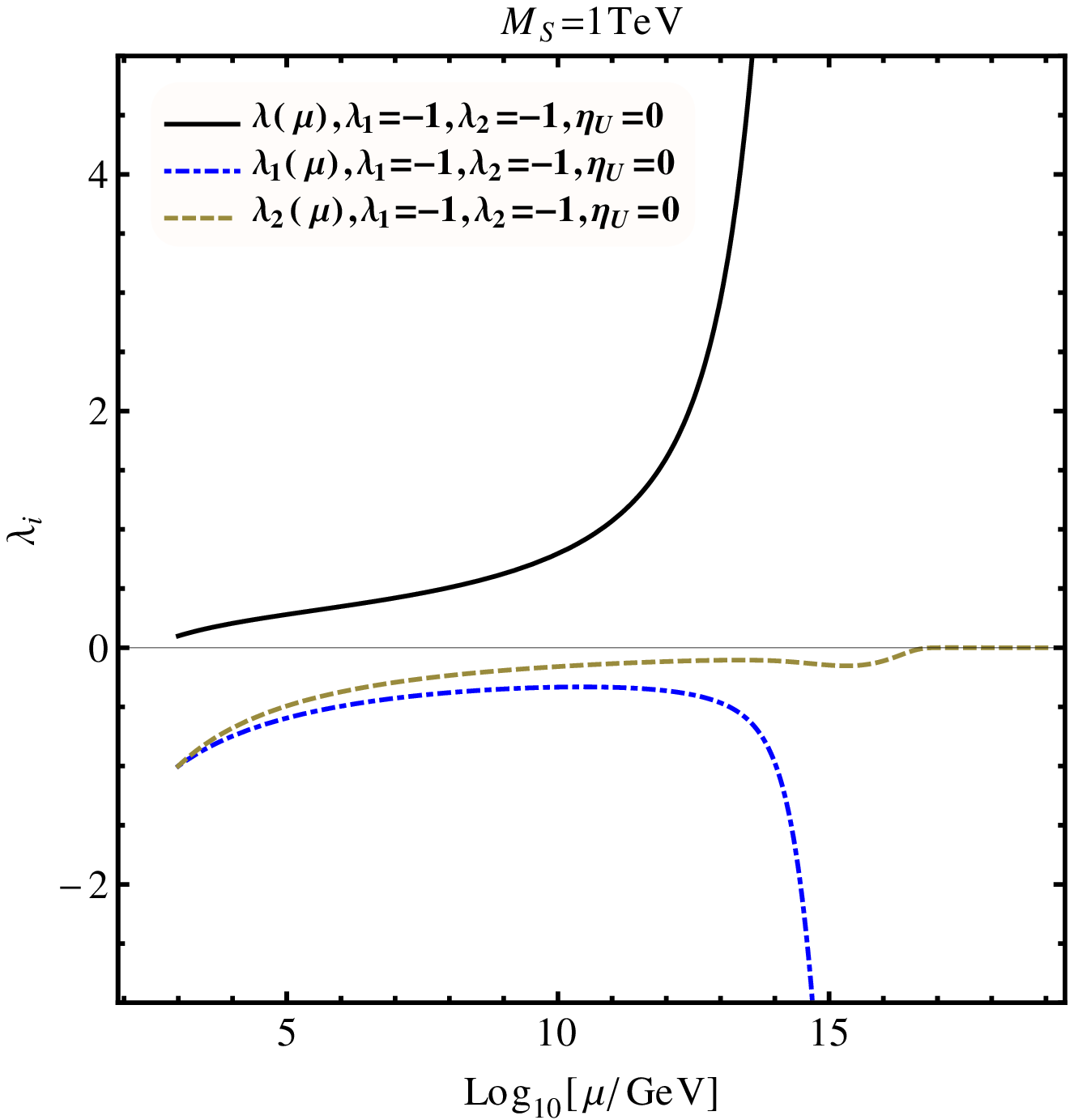}
\caption{
\label{fig:lambdairunning}Illustration of running behaviors of $\lambda_i$ and $\lambda$.}
\end{figure}

Finally we study the sector with $\lambda_{8,9,11}$ responsible for high energy scattering $SS \to SS$ at tree level in the custodial symmetry limit. The corresponding RGE are given by
\begin{eqnarray}
\frac{d\lambda_8}{d\ln \mu}&=&  \frac{1}{144\pi^2}(180\lambda_{8}^{2}+234\lambda_{8}(\lambda_{9}+\lambda_{11})+54\lambda_{9}^{2}+67\lambda_{11}^{2})\nonumber\\
\frac{d\lambda_9}{d\ln \mu}&=& \frac{1}{144\pi^2}(144\lambda_{9}^{2}+54\lambda_{8}\lambda_{9}+78\lambda_{9}\lambda_{11}+\lambda_{11}^{2})\nonumber\\
\frac{d\lambda_{11}}{d\ln \mu}&=& \frac{1}{32\pi^2}( 13\lambda_{11}^{2}+12\lambda_{8}\lambda_{11}+12\lambda_{9}\lambda_{11})
\end{eqnarray}
This set of coupled equations can be solved numerically and we obtain results that are qualitatively similar to the ones for $\lambda_{1,2}$. As we require the unitarity constraints to be satisfied to higher energies, the allowed parameter space for positive values of $\lambda_{8,9,11}$ shrinks and they are constrained to be small. In Fig.~\ref{fig:figl18-l11} we show our numerical solution for the allowed regions in $\lambda_{8,9,11}$ (at 1~TeV) that satisfy the unitarity constraints Eqs.~\ref{octetb1},~\ref{octetb2}, and~\ref{octetb3} up to $10^{10}$~GeV by taking two of them to be non-zero at a time as the blue region. 

\begin{figure}[thb]
\includegraphics[width=0.3\textwidth]{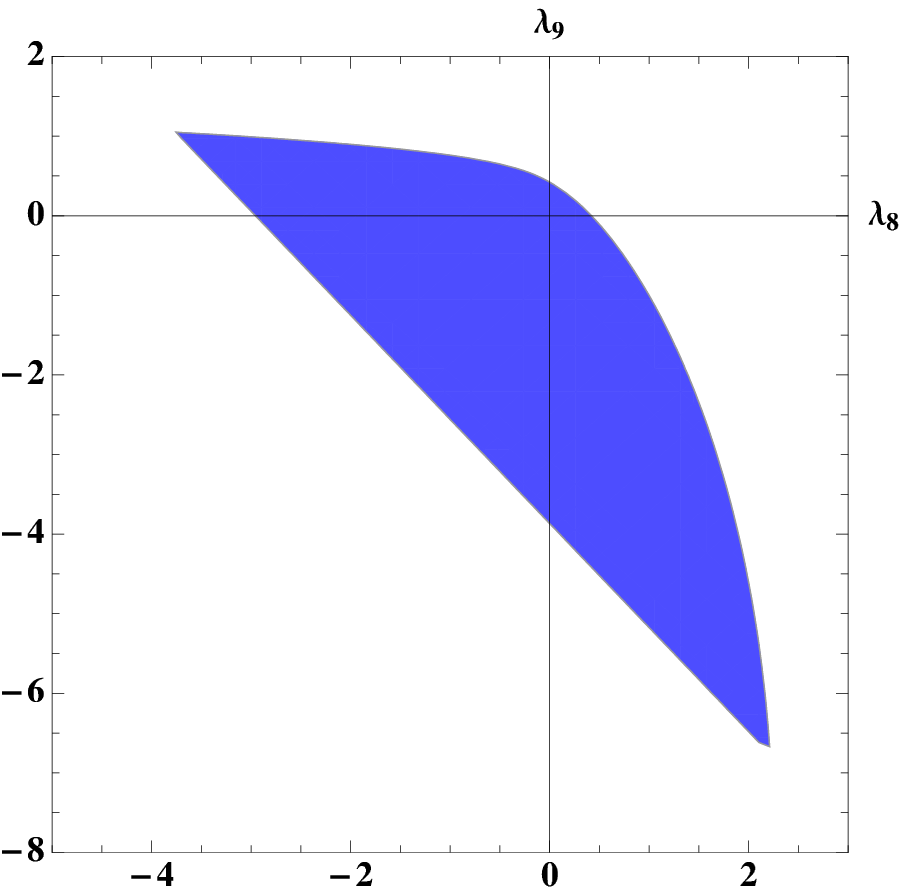}
\includegraphics[width=0.305\textwidth]{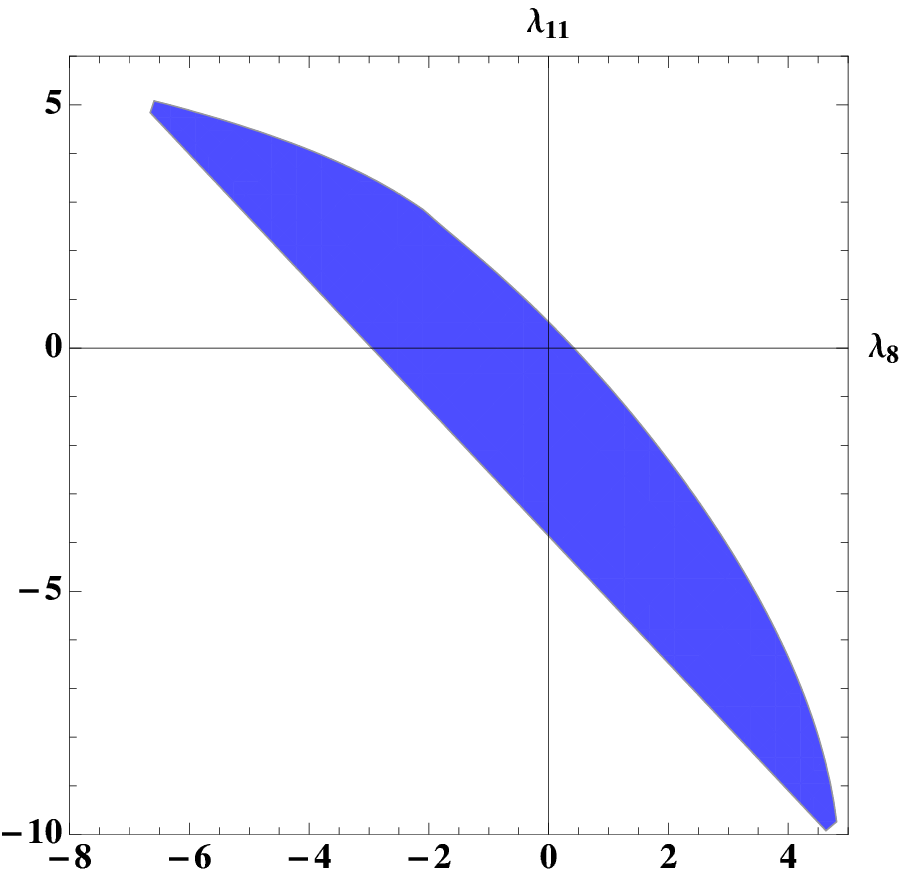}
\includegraphics[width=0.3\textwidth]{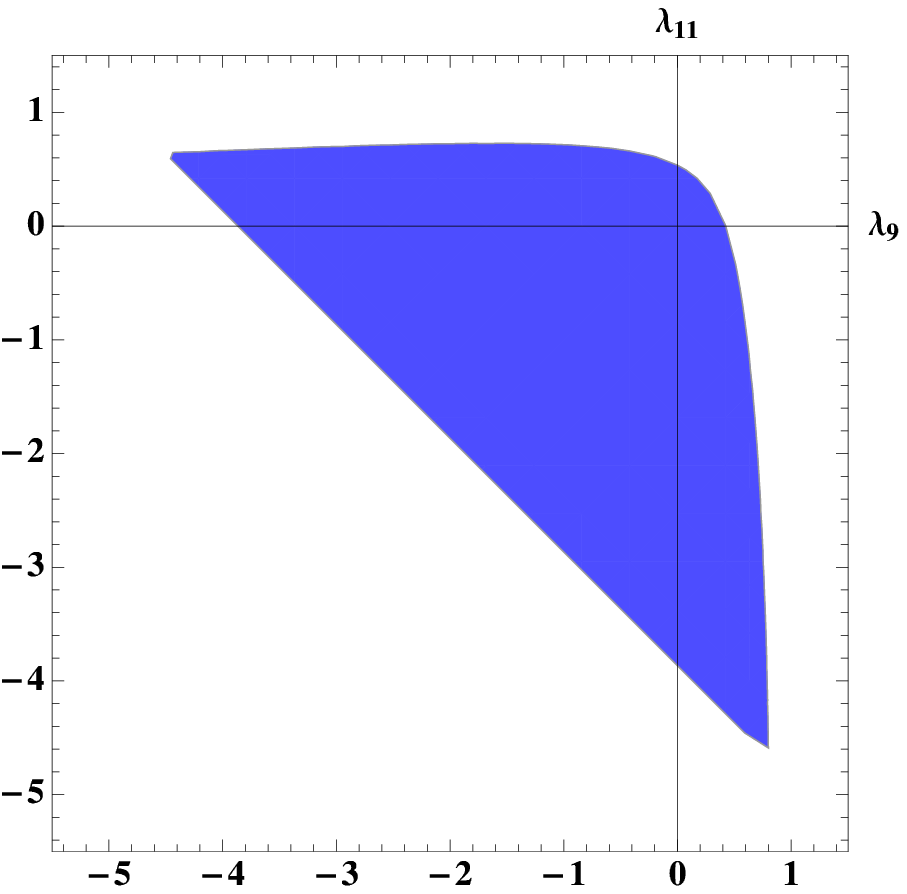}
\caption{
\label{fig:figl18-l11}Allowed regions in $\lambda_{8,9,11}$ (at 1~TeV) that satisfy the unitarity constraints Eqs.~\ref{octetb1},~\ref{octetb2},~\ref{octetb3} up to $10^{10}$~GeV by taking two of them to be non-zero at a time shown in the blue region.}
\end{figure}

\section{Numerical values for color-octet couplings used in the literature}

In this section we briefly review some of the phenomenological applications of this color-octet model that have appeared in the literature with emphasis on the numerical values used for the different couplings. 

In their original paper Manohar and Wise \cite{Manohar:2006ga} point out that the values $\lambda_1 =4$ and $\lambda_2 =1$ for $m_S=750$~GeV can increase the Higgs production rate by a factor of 2 while decreasing $H\to \gamma\gamma$ by 10\% and being consistent with constraints from the $S$ parameter.

In \cite{Gresham:2007ri}, it is found that the contributions of the model to $R_b$ imply that for masses in the $m_S\sim 1\ (4)$~TeV region, $|\eta_U| \lsim 1.8\ (5)$ to stay within 2$\sigma$ of the measured value.  They also compute the singlet color octet production rate at LHC  assuming $\lambda_{4,5}\sim1$ and $|\eta_U|=1$.

In \cite{Gerbush:2007fe}, the minimum values of $|\lambda_2\pm2\lambda_3|$ and of $|\lambda_3|$ that allow for two body decays of the color octet states involving $W^\pm$ are calculated. For a 
${\rm min}\{m_{S^\pm},m_{S_{i,r}}\}$ in the range $200-600$~GeV, $|\lambda_2\pm2\lambda_3|_{\rm min} \sim (2-6)$ for example. In addition they also consider the running of the couplings $\lambda_i$ and find constraints by requiring that they remain perturbative up to some  scale which they use to study color octet phenomenology at LHC. We disagree with their beta-functions as given in their appendix so the constraints that we obtain are different.

In \cite{Burgess:2009wm}, the mass splittings between $M_+, M_R, M_I$ are studied imposing constraints from electroweak precision data and allowed regions are presented for $\lambda_{1,2,3}$ as large as 10.  Requiring that there be no Landau pole in the running of $\lambda$ up to 10~TeV results in $\lambda_1\lsim 1.3$ and $\sqrt{\lambda_1^2+\lambda_2^2}\lsim 2.2$. They also point out the electric dipole moment of the neutron for light masses $m_S = 100$~GeV implies that ${\rm Im}(\eta_U^\star\eta_D^\star)<0.1$. Finally they argue that  $\lambda_{4,5}<10$ to remain in the perturbative regime.

Several studies use this model to modify jet production. For example, in \cite{Carpenter:2011yj,Enkhbat:2011qz}, this model is studied in connection with the CDF diet anomaly. The production cross section of $S_R^0$ is calculated as function of $\lambda_4+\lambda_5$, ranging from 0 to 30. For the benchmark point of $M_R=245$~GeV, $M_I=190$~GeV, $M_{\pm}=150$~GeV, $\lambda_4+\lambda_5=18$ is needed to explain the CDF anomaly. In Ref.~\cite{Arnold:2011ra} several multi jet process are studied with $\lambda_{4,5}\sim1$ and $\lambda_{1,2,3}\sim 1/2$.

Several papers have studied this model in relation to possible enhancements/suppressions of the Higgs coupling to gluons or photons. In Ref.~\cite{He:2011ti} it is argued that this model can reduce the Higgs production cross section at LHC to make it compatible with a fourth generation. Saturating the $S$ parameter with $\lambda_2$ results in the constraint for $m_S\sim1$~TeV, $-48 \lsim \lambda_2 \lsim 6$. Using this constraint on $\lambda_2$ it is found that $\lambda_1 \sim -8$ would be needed to halve the SM4 $hgg$ coupling for masses $m_S \sim 2v$. A similar argument for reducing the Higgs production cross section in the SM was presented in Ref.~\cite{Dobrescu:2011aa} and one for hiding heavier Higgs bosons in Ref.~\cite{Bai:2011aa}. Ref.~\cite{Dorsner:2012pp} uses similar models to enhance the $H\to\gamma\gamma$ rate and Ref.~\cite{Kribs:2012kz} to enhance di-Higgs production.

In Ref.~\cite{He:2011ws} it is pointed out that this model can result in large CP violation in top pair production at the LHC. The parameter range considered was $\lambda_{4,5}\lsim 8$ and large phases in $\lambda_4$ and $\eta_U$. It was also found that for color octet resonances with masses in the $500-1000$~GeV range one needed $\eta_U \sim 3$ to have them stand out over QCD background.

In \cite{Cacciapaglia:2012wb} the $hgg$ and $h\gamma\gamma$ couplings as fit from LHC data are compared with sample BSM scenarios. Although no attempt is made to constrain the couplings 
of the color octet model, it is shown that it is consistent with data using $m_s = 750$~GeV for values  $\lambda_1=4, \lambda_2=1$. 

In Ref.~\cite{Reece:2012gi} the possibility of inverting the sign of the $hgg$ coupling with additional color octets is studied and the allowed parameter space has $\lambda_O \gsim 4$ where $\lambda_O$ corresponds to $\lambda_8$ in Eq.~\ref{potential}.

Finally, in the very recent Ref.~\cite{Cao:2013wqa} the model is considered in connection to LHC Higgs data. They examine the constraints imposed on the model by unitarity numerically and find that for $m_h = 125.5$~GeV $|\lambda_{1,2}|\lsim 35$ and $|\lambda_3|\lsim 18$, worse than our limits by factors between 2 and 4.

Color octet scalars also appear in specific models considered recently \cite{Bertolini:2013vta}.

\section{Summary and Conclusions}

In this paper we have examined the theoretical constraints on the color-octet extension of the standard model scalar sector. We first required perturbative unitarity for two-to-two scalar scattering amplitudes to obtain general bounds for the parameters in the scalar potential. We then considered the renormalization group equations for the couplings and used this to provide constraints from the requirement of a stable vacuum up to high energy. An amusing fact is that for octet masses near a TeV, even with vanishing couplings to quarks and the higgs boson, the electroweak vacuum can be stabilized up to the Planck scale. Finally we considered improvement of our unitarity constraints by requiring they be satisfied up to high energy scales. This consideration further constrains the values allowed. Finally we reviewed some of the phenomenological studies that have used this model in the literature and found that many of them stray outside the theoretical constraints found here. Our results should prove useful for future phenomenology of this model. 

\begin{acknowledgments}

The work of XGH and YT was supported in part by NSC of ROC, and XGH was also supported in part by NNSF(grant No:11175115) and Shanghai science and technology commission (grant No: 11DZ2260700) of PRC. The work of GV and HP was supported in part by DOE under contract number DE-FG02-01ER41155. 

\end{acknowledgments}

\appendix

\section{Partial wave amplitudes in the general case without custodial symmetry}

We begin by displaying results for $hh \to SS$ partial wave amplitudes in the high energy limit in Table~\ref{t:hhss}.
\begin{table}[htdp]
\caption{$J=0$ partial wave amplitudes for the scattering of the  different neutral combinations of $hh\to SS$ in a color singlet state.}
\begin{center}
\begin{tabular}{|c|c|}
\hline
channel & $8\sqrt{2}\pi\ a^0_{J=0}$ \\ \hline
$w^+w^-\to S^+S^-$ & $\lambda_1+\lambda_2$ \\  \hline
$w^+w^-\to S_iS_i$ \ =\ $w^+w^-\to S_rS_r$\ =\ $zz\to S^+S^-$ \ =\  $hh\to S^+S^-$& $\lambda_1$ \\  \hline
$zz\to S_iS_i$\  =\ $hh \to S_rS_r$ & $\lambda_1+\lambda_2+2\lambda_3$ \\  \hline
$zz\to S_rS_r$\ =\ $hh \to S_iS_i$ & $\lambda_1+\lambda_2-2\lambda_3$ \\  \hline
\end{tabular}
\end{center}
\label{t:hhss}
\end{table}%

Inspection of the partial wave amplitudes presented in Table~\ref{t:hhss} reveals that Eq.~\ref{vvss} is obtained by adding all the channels (with a suitable normalization). Other possible combinations  include
\begin{eqnarray}
a((zz+hh)/2 \to (S_iS_i + S_rS_r)/2)_{J=0}^0 &=& \frac{(\lambda_1+\lambda_2)}{8\sqrt{2}\pi} \nonumber \\
a((zz-hh)/2 \to S_iS_i)_{J=0}^0 &=& \frac{\lambda_3}{4\sqrt{2}\pi}.
\label{morehhss}
\end{eqnarray}
From the last of these conditions we obtain Eq.~\ref{l3bound}, and from the second row in Table~\ref{t:hhss} we obtain $\lambda_1\lsim 18$.

Next we consider the scattering of $SS\to SS$ to constrain $\lambda_{4-11}$ and display the  results for $J=0$ partial wave amplitudes in Table~\ref{t:ssss}.
\begin{table}[htdp]
\caption{$J=0$ partial wave amplitudes for the scattering of the  different neutral combinations of $SS\to SS$ in a color singlet state.}
\begin{center}
\begin{tabular}{|c|c|}
\hline
channel & $96\pi\ a^0_{J=0}$ \\ \hline
$S^+S^-\to S^+S^-$ & $6 \lambda_{10}+7 \lambda_{11}+16 \lambda_{6}+16 \lambda_{7}+27 \lambda_{8}+27 \lambda_{9}$ \\  \hline
$S_iS_i\to S_iS_i$\ = \ $S_rS_r\to S_rS_r$ & $30 \lambda_{10}+15 \lambda_{11}+15 \lambda_{6}+15 \lambda_{7}+30 \lambda_{8}+30 \lambda_{9}$ \\  \hline
$S_iS_i\to S^+S^-$\ = \ $S_rS_r\to S^+S^-$ & $3 \lambda_{10}+8 \lambda_{11}+8 \lambda_{6}+8 \lambda_{7}+24 \lambda_{8}+3 \lambda_{9}$ \\  \hline
$S_iS_i\to S_rS_r$ & $3 \lambda_{10}-\lambda_{11}+17 \lambda_{6}+17 \lambda_{7}+24 \lambda_{8}+3 \lambda_{9}$ \\  \hline
\end{tabular}
\end{center}
\label{t:ssss}
\end{table}%
In the custodial symmetry limit the corresponding results are shown in Table~\ref{t:ssssc}.
\begin{table}[htdp]
\caption{$J=0$ partial wave amplitudes for the scattering of the  different neutral combinations of $SS\to SS$ in a color singlet state in the custodial symmetry limit.}
\begin{center}
\begin{tabular}{|c|c|}
\hline
channel & $96\pi\ a^0_{J=0}$ \\ \hline
$S^+S^-\to S^+S^-$ & $23 \lambda_{11}+27 \lambda_{8}+33 \lambda_{9}$ \\  \hline
$S_iS_i\to S_iS_i$\ = \ $S_rS_r\to S_rS_r$ & $30 \lambda_{11}+30 \lambda_{8}+60 \lambda_{9}$ \\  \hline
$S_iS_i\to S^+S^-$\ = \ $S_rS_r\to S^+S^-$\ = \ $S_iS_i\to S_rS_r$ & $16 \lambda_{11}+24 \lambda_{8}+6 \lambda_{9}$ \\  \hline
\end{tabular}
\end{center}
\label{t:ssssc}
\end{table}%

Weaker bounds arise from considering color octet (or 27) channels but they can be useful to place separate constraints on particular couplings. For illustration we include one such channel in the antisymmetric color octet
\begin{eqnarray}
a^0_{(+-)0}&=&a(\frac{1}{\sqrt{2}}(S^+S^--S^-S^+)\to\frac{1}{\sqrt{2}}(S^+S^--S^-S^+))_{J=0}^{8A}\nonumber \\
&=&\frac{1}{32\pi}(2\lambda_8-2\lambda_9+3\lambda_{11})+\frac{3}{32\pi}\frac{v^2}{(s-m_s^2)}\lambda_4^2\sin^2(\phi_4)
\nonumber \\
&-&\frac{1}{32\pi}\frac{v^2}{(s-m_s^2)}\lambda_1^2
\log\left(\frac{s-4m_s^2+m_H^2}{m_H^2}\right) \nonumber \\
&-&\frac{1}{192\pi}\frac{v^2}{(s-4m_s^2)}\lambda_4^2(7-2\cos(2\phi_4))\log\left(\frac{s-3m_s^2}{m_s^2}\right)
\label{ssss18}
\end{eqnarray}
from which one finds
\begin{eqnarray}
\left|2\lambda_8-2\lambda_9+3\lambda_{11}\right| \lsim 16\pi.
\label{octetb3}
\end{eqnarray}

\section{$\beta$ functions without custodial symmetry}
\label{fullbetas}
Using the techniques in ref.~\cite{Cheng:1973nv,Branco:2011iw}, one can easily obtain the relevent $\beta$ functions. Here we give the color octet contributions to the $\beta$ functions for $\lambda$s in the general case without custodial symmetry,
\begin{eqnarray*}
(16\pi^{2})\beta_{\lambda_{1}} & = &  2 \lambda_1^2 +\lambda_2^2 + 4\lambda_3^2 + 4 \lambda(3 \lambda_1+\lambda_2) \\
&&{} + \frac{1}{3}(7\lambda_{4}\lambda_{4}^{*}-2\lambda_{4}^{*}\lambda_{5}-2\lambda_{5}^{*}\lambda_{4} + 7\lambda_{5}\lambda_{5}^{*})\\
&&{} +\lambda_{1}(8\lambda_{6}+8\lambda_{7}+17\lambda_{8}+10\lambda_{9}+3\lambda_{10}+5\lambda_{11})\\
&&{}+\frac{1}{3}\lambda_{2}(8\lambda_{6}+8\lambda_{7}+24\lambda_{8}+3\lambda_{9}+3\lambda_{10}+8\lambda_{11}),
\end{eqnarray*}
\begin{eqnarray*}
(16\pi^{2})\beta_{\lambda_{2}} & = & 2 \lambda_2^2 + 4\lambda_1\lambda_2+ 16\lambda_3^2 + 4 \lambda \lambda_2\\
&&{}+\frac{1}{6}(8\lambda_{4}\lambda_{4}^{*}+17\lambda_{4}^{*}\lambda_{5}+17\lambda_{5}^{*}\lambda_{4}+8\lambda_{5}\lambda_{5}^{*})\\
&&{}+\frac{1}{6}\lambda_{2}(16\lambda_{6}+16\lambda_{7}+6\lambda_{8}+48\lambda_{9}+6\lambda_{10}-2\lambda_{11}),
\end{eqnarray*}
\begin{eqnarray*}
(16\pi^{2})\beta_{\lambda_{3}} & = & 2 \lambda_3(2\lambda + 2\lambda_1 + 3\lambda_2)+\frac{1}{12}(17\lambda_{4}^{2}+16\lambda_{4}\lambda_{5}+17\lambda_{5}^{2})\\
&&{} +\frac{1}{3}\lambda_{3}(-\lambda_{6}-\lambda_{7}+3\lambda_{8}+3\lambda_{9}+24\lambda_{10}+8\lambda_{11}),
\end{eqnarray*}
\begin{eqnarray*}
(16\pi^{2})\beta_{\lambda_{4}} & = & 8\lambda_{4}^{*}\lambda_{3}+2\lambda_{3}\lambda_{5}^{*}+\lambda_{5}(2\lambda_{2}-\lambda_{7}+2\lambda_{9}+4\lambda_{10}+\lambda_{11})\\
&&+\lambda_{4}(3\lambda_{1}+2\lambda_{2}+6\lambda_{6}+2\lambda_{7}+3\lambda_{8}+2\lambda_{9} + \lambda_{10}+\lambda_{11}),
\end{eqnarray*}
\begin{eqnarray*}
(16\pi^{2})\beta_{\lambda_{5}} & = & 2\lambda_{3}\lambda_{4}^{*}+8\lambda_{3}\lambda_{5}^{*}+\lambda_{4}(2\lambda_{2}-\lambda_{6}+2\lambda_{9}+4\lambda_{10}+\lambda_{11})\\
 &  & +\lambda_{5}(3\lambda_{1}+2\lambda_{2}+6\lambda_{6}+2\lambda_{7}+3\lambda_{8}+2\lambda_{9}+\lambda_{10}+\lambda_{11}),
\end{eqnarray*}
\begin{eqnarray*}
(16\pi^{2})\beta_{\lambda_{6}} & = & 3\lambda_{4}\lambda_{4}^{*}+7\lambda_{6}^{2}+\lambda_{6}(6\lambda_{7}+6\lambda_{8}+4\lambda_{9}-\lambda_{10}-2\lambda_{11})\\
 &  & +\lambda_{7}(4\lambda_{9}-\lambda_{10})-2\lambda_{9}\lambda_{11}+2\lambda_{10}\lambda_{11}+\lambda_{11}^{2},
\end{eqnarray*}
\begin{eqnarray*}
(16\pi^{2})\beta_{\lambda_{7}} & = & 3\lambda_{5}\lambda_{5}^{*}+7\lambda_{7}^{2}+\lambda_{7}(6\lambda_{6}+6\lambda_{8}+4\lambda_{9}-\lambda_{10}-2\lambda_{11})\\
 &  & +\lambda_{6}(4\lambda_{9}-\lambda_{10})-2\lambda_{9}\lambda_{11}+2\lambda_{10}\lambda_{11}+\lambda_{11}^{2},
\end{eqnarray*}
\begin{eqnarray*}
(16\pi^{2})\beta_{\lambda_{8}} & = & 20\lambda_{8}^{2}+\frac{1}{18}\lambda_{8}(288\lambda_{6}+288\lambda_{7}+360\lambda_{9}+108\lambda_{10}+180\lambda_{11})\\
 &  & +\frac{1}{18}[36\lambda_{1}^{2}+36\lambda_{1}\lambda_{2}-24\lambda_{4}\lambda_{4}^{*}-6\lambda_{4}^{*}\lambda_{5}-6\lambda_{5}^{*}\lambda_{4}-24\lambda_{5}\lambda_{5}^{*}\\
 &  & +62\lambda_{6}^{2}+64\lambda_{6}\lambda_{7}+62\lambda_{7}^{2}+96\lambda_{9}(\lambda_{6}+\lambda_{7})+18\lambda_{10}(\lambda_{6}+\lambda_{7})+58\lambda_{11}(\lambda_{6}+\lambda_{7})\\
 &  & +54\lambda_{9}^{2}+36\lambda_{9}\lambda_{10}+132\lambda_{9}\lambda_{11}+18\lambda_{10}^{2}+18\lambda_{10}\lambda_{11}+29\lambda_{11}^{2}],
\end{eqnarray*}
\begin{eqnarray*}
(16\pi^{2})\beta_{\lambda_{9}} & = & \lambda_{2}^{2}-\frac{1}{3}(\lambda_{4}\lambda_{4}^{*}-2\lambda_{4}^{*}\lambda_{5}-2\lambda_{5}^{*}\lambda_{4}+\lambda_{5}\lambda_{5}^{*})+10\lambda_{9}^{2}\\
 &  & +\lambda_{10}(\lambda_{6}+\lambda_{7}+\lambda_{11})+\lambda_{9}[\frac{4}{3}(4\lambda_{6}+4\lambda_{7}+\lambda_{11})+2\lambda_{10}+6\lambda_{9}]\\
 &  & +4\lambda_{10}^{2}+\frac{1}{9}(\lambda_{6}^{2}+\lambda_{7}^{2}-4\lambda_{11}(\lambda_{6}+\lambda_{7})-2\lambda_{11}^{2})+\frac{26}{9}\lambda_{6}\lambda_{7},
\end{eqnarray*}
\begin{eqnarray*}
(16\pi^{2})\beta_{\lambda_{10}} & = & 4\lambda_{3}^{2}-\frac{1}{3}(\lambda_{4}\lambda_{4}^{*}-2\lambda_{4}^{*}\lambda_{5}-2\lambda_{5}^{*}\lambda_{4}+\lambda_{5}\lambda_{5}^{*})\\
 &  & +\lambda_{10}[\frac{1}{3}(\lambda_{6}+\lambda_{7}+19\lambda_{11})+8\lambda_{9}+6\lambda_{8}]+2\lambda_{9}\lambda_{11}+8\lambda_{10}^{2}\\
 &  & +\frac{1}{9}(\lambda_{6}^{2}+\lambda_{7}^{2}-4\lambda_{11}(\lambda_{6}+\lambda_{7})+7\lambda_{11}^{2})-\frac{10}{9}\lambda_{6}\lambda_{7},
\end{eqnarray*}
and finally
\begin{eqnarray*}
(16\pi^{2})\beta_{\lambda_{11}} & = & \frac{1}{2}\lambda_{11}^{2}+3\lambda_{4}\lambda_{4}^{*}+3\lambda_{5}\lambda_{5}^{*}-2(\lambda_{6}^{2}+\lambda_{7}^{2})+6\lambda_{10}(\lambda_{6}+\lambda_{7})+7\lambda_{11}(\lambda_{6}+\lambda_{7}+\lambda_{8}). 
\end{eqnarray*}

\end{document}